\documentclass[12pt]{article}
\input psfig.sty

\def\xbe7{$^7Be$}
\def\be7pg{$^7Be(p,\gamma)^8B$}
\def\b8{$^8B$}
\def\n16{$^{16}N$}
\def\xo16{$^{16}O$}
\def\co2{$CO_2$}
\def\c12ag{$^{12}C(\alpha,\gamma)^{16}O$}
\def\go16{$^{16}O(\gamma,\alpha)^{12}C$}
\def\sE1{$S_{E1}(300)$}
\def\xsE2{$S_{E2}(300)$}

\begin{document}

\title{
 KeV Astrophysics With GeV Beams; \\
  Blazing a New Trail on the Summitt of Nuclear Astrophysics 
\footnote{Work Supported by USDOE Grant No. DE-FG02-94ER40870.}}

\author
{Moshe Gai \\
 Laboratory for Nuclear Science, Dept. of Physics, U3046, \\ 
University of Connecticut, 
    2152 Hillside Rd., Storrs, CT 06269-3046, USA \\ 
   (gai@uconn.edu, http://www.phys.uconn.edu)}

\maketitle

\begin{abstract}

GeV beams of light ions and electrons are used for creating a 
high flux of real and virtual photons, with which some  
problems in Nuclear Astrophysics are studied.  GeV \b8 beams 
are used to study the Coulomb 
dissociation of $^8B$ and thus the \be7pg reaction.  
This reaction is one of the major source of uncertainties 
in estimating the \b8 solar neutrino flux and a critical 
input for calculating the \b8 Solar neutrino flux. 
The Coulomb dissociation of \b8 appears to provide a viable method 
for measuring the \be7pg reaction rate, with a weighted average of 
the RIKEN1, RIKEN2, GSI1 and MSU published results of 
$ S_{17}(0) = 18.9 \pm 1.0 $ eV-b. This result however does 
not include a theoretical error estimated to be $\pm 10 \%$. 
GeV electron beams on the other hand, are used to create a high 
flux of real and virtual photons at TUNL-HIGS and MIT-Bates, 
respectively, and we discuss two new proposals 
to study the \c12ag reaction with real and virtual 
photons. The \c12ag reaction is 
essential for understanding Type II and Type Ia supernova. 
It is concluded that virtual and real photons produced by 
GeV light ions and electron beams are useful for studying 
some problems in Nuclear Astrophysics.

\end{abstract}

\section{Introduction: The \be7pg Reaction at Low Energies}

The solar neutrino problem \cite{Book,Phs96} may allow for new
break through in neutrino physics and 
the standard model of particle physics.
The precise knowledge of the astrophysical $S_{17}$-factor 
[$= \ \sigma _{17} \times E \times exp(2 \pi \eta )$ where 
$\eta$ is the Sommerfeld parameter 
$= \ Z_1 Z_2 {\alpha \over \beta} $), is crucial 
for interpreting terrestrial measurements of the solar neutrino
flux~\cite{Ade98}.
This is particularly true for the interpretation of results from the
Homestake, Kamiokande, SuperKamiokande and SNO experiments
~\cite{Phs96} which
measured high energy solar neutrinos mainly or solely from $^8$B decay.
In Fig. 1 we show the world data including our new GSI measurement of
$S_{17}$ \cite{Iw99}.  It is clear from this
figure that if we are to quote $S_{17}(0)$
with an accuracy of $\pm 5\%$, many more experiments using different
methods are required.

\centerline{\psfig{figure=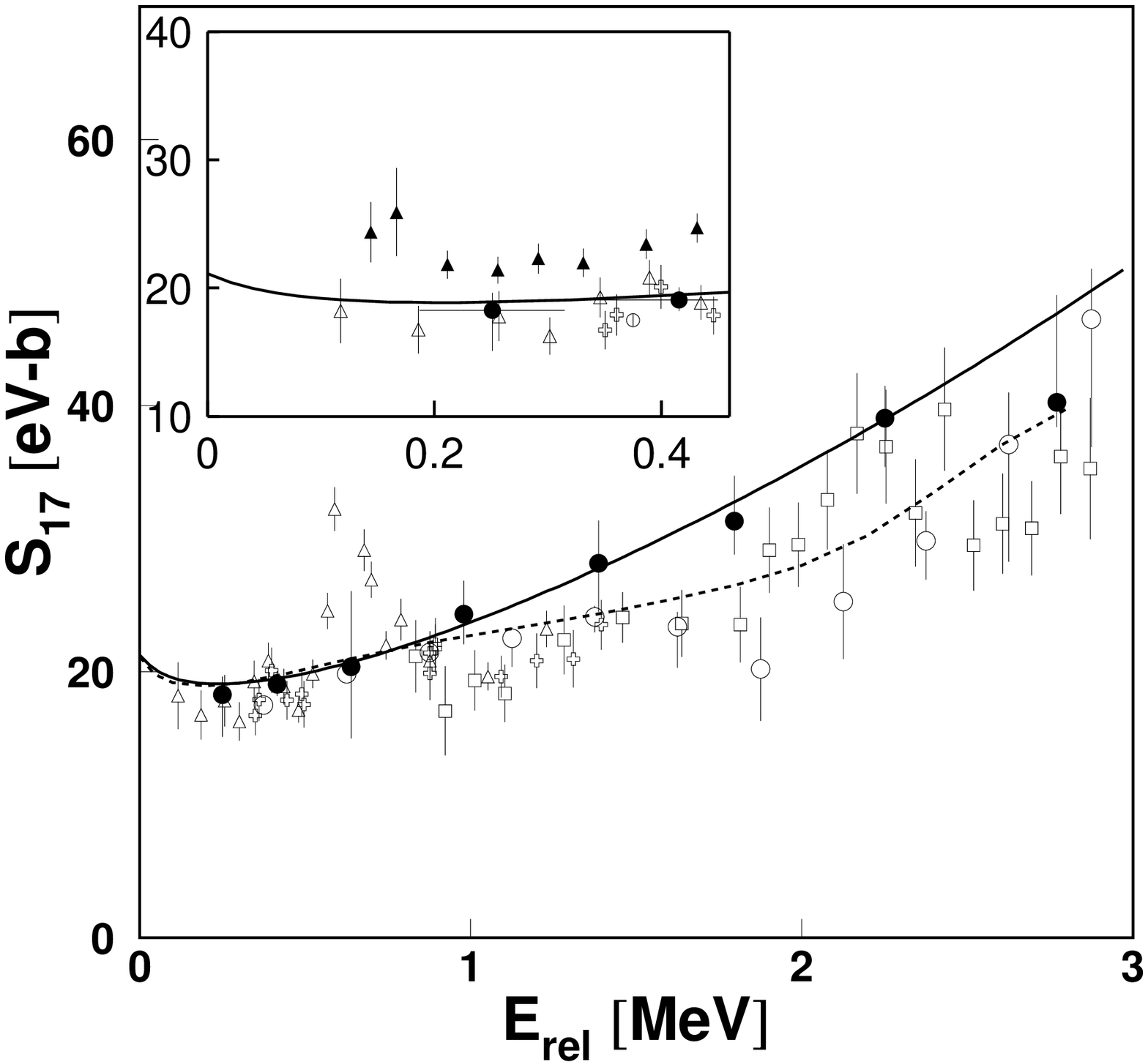,height=4in}}

\hspace{0.5in} \underline{Fig. 1:}  World data on $S_{17}$ as reported
       by our GSI collabortion \cite{Iw99}.

\section{The Coulomb Dissociation of $^8B$ and the \be7pg Reaction
     at Low Energies}

The Coulomb Dissociation (CD) \cite{Bau86,Be88} is a
Primakoff \cite{Pr51} process that could be viewed in first order as the time
reverse of the radiative capture reaction. The large relative velocities 
of light ion beams create a large flux of virtual photons and 
instead of studying for example the fusion of a proton 
plus a nucleus (A-1), one studies the
disintegration of the final nucleus (A) in the Coulomb field, to a proton
plus the (A-1) nucleus.  The reaction is made possible by the absorption of
a virtual photon from the field of a high Z nucleus such as $^{208}Pb$.  
In this case since $\pi/k^2$  for a photon is approximately 
100-1000 times larger than that
of a particle beam, the small cross section is enhanced.  The large virtual
photon flux (typically 100-1000 photons per collision) also gives rise to
enhancement of the cross section.  Our understanding of the Coulomb
dissociation process \cite{Bau86,Be88} allows us to extract the inverse
nuclear process even when it is very small.  However in Coulomb
dissociation since $\alpha Z$  approaches unity (unlike the case in electron
scattering), higher order Coulomb effects (Coulomb post acceleration) may
be non-negligible and they need to be understood
\cite{Ber94,Typ94}.  The success of CD experiments \cite{Ga95} is in
fact contingent on understanding such effects and
designing the kinematical conditions so as to minimize such effects.

Hence the Coulomb dissociation process has to be measured with great care
with kinematical conditions carefully adjusted so as to minimize nuclear
interactions (i.e. distance of closest approach considerably larger then 20
fm, or very small forward angles scattering), and measurements must be
carried out at high enough energies (many tens of MeV/u) so as to maximize
the virtual photon flux.

\subsection{The GSI Results at 254 A MeV}

An experiment to measure the Coulomb dissociation of $^8$B at a
higher energy of 254~$A$~MeV was performed at GSI \cite{Iw99}.
The present experimental conditions have several advantages:
(i) forward focusing allows us to use the magnetic
spectrometer KaoS~\cite{Senger93} at GSI for
a kinematically complete measurement with high detection
efficiency over a wider range of the p-$^7$Be relative energy;
(ii) because of the smaller influence of straggling on the
experimental resolution at the higher energy, a thicker target
can be used for compensating the weaker beam intensity, (iii)
effects that obscure the contribution of E1 multipolarity to
the Coulomb dissociation like E2 admixtures and higher-order
contributions are reduced~\cite{Ber94,Typ94}. The contribution
of M1 multipolarity is expected to be enhanced at the higher energy,
but this allows to observe the M1 resonance peak and determine its
$\gamma$ width.

A $^8$B beam was produced by fragmentation of a 350~$A$~MeV $^{12}$C
beam from the SIS synchrotron at GSI that impinged on a beryllium
target with a thickness of 8.01 g/cm$^2$.
The beam was isotopically separated by the fragment separator
(FRS)~\cite{Geissel92}
by using an aluminum degrader with a thickness of 1.46 g/cm$^2$
with a wedge angle of 3 mrad.
The beam was transported to the standard target-position of
the spectrometer KaoS~\cite{Senger93}.
The average beam energy of $^8$B in front of the breakup
target was 254.5~$A$~MeV,
a typical $^8$B intensity was 10$^4$ /spill (7s/spill).
Beam-particle identification was achieved event by event with
the TOF-$\Delta E$ method by using a beam-line plastic scintillator
with a thickness of 5 mm placed
68 m upstream from the target and a large-area scintillator wall
discussed later placed close to the focal plane of KaoS.
About 20 \% of the beam particles were $^7$Be, which could
however unambiguously be
discriminated from breakup $^7$Be particles by their time of flight.

An enriched $^{208}$Pb target with a thickness of
199.7 ($\pm$ 0.2) mg/cm$^2$ was placed at the entrance of KaoS.
The average energy at the center of the target amounted
to 254.0~$A$~MeV. The reaction products, $^7$Be and proton,
were analyzed by the spectrometer
which has a large momentum acceptance of $\Delta p/p \approx 50$~\% and
an angular acceptance of 140 and 280~mrad in horizontal and vertical
directions, respectively.
For scattering-angle measurement or track reconstruction of
the two reaction products,
two pairs of silicon micro-strip detectors were installed
at about 14 and 31~cm downstream from the target, respectively,
measuring either x- or y-position of the products before
entering the KaoS magnets.
Each strip detector had a thickness of 300~$\mu$m, an active
area of 56 $\times$ 56 mm$^2$, and a strip pitch of 0.1 mm.

The measured complete kinematics of the breakup products
allowed us to reconstruct the p-$^7$Be relative energy and
the scattering angle $\theta_8$ of
the center-of-mass of proton and $^7$Be (excited $^8$B) with
respect to the incoming beam from the individual momenta and
angles of the breakup products.

To evaluate the response of the detector system, Monte-Carlo
simulations were performed using the code GEANT\cite{GEANT}.
The simulations took into account
the measured $^8$B beam spread in energy, angle, and position
at the target, as well as the influence of angular and energy
straggling and energy loss in the layers of matter.
Losses of the products due to limited detector sizes were also accounted for.
Further corrections in the simulation are due to the feeding of
the excited state at 429~keV in $^7$Be.
We used the result by Kikuchi {\em et al.}~\cite{Ki97} who
measured the $\gamma$-decay in coincidence with Coulomb
dissociation of $^8$B at 51.9~$A$~MeV.

The Monte Carlo simulations yielded relative-energy resolutions
from the energy and angular resolutions of the detection system
to be 0.11 and 0.22~MeV (1$\sigma$) at $E_{\rm rel}=0.6$ and
1.8 MeV, respectively.
The total efficiency calculated by the simulation was
found to be larger than 50\% at $E_{\rm rel}= 0-2.5$~MeV.
due to the large acceptance of KaoS.
The experimental angular distributions are well reproduced by the
simulation using only E1 multipolarity, in line with the
results of Kikuchi {\it et al.}~\cite{Ki97} and Gai and 
Bertulani \cite{Ga95}.

To study our sensitivity to a possible E2 contribution,
we have added the simulated E2
angular distribution from the E2 Coulomb dissociation
cross section calculated by Bertulani and Gai \cite{Ber98}.
Note that nuclear breakup effects are also included in
the calculation. By fitting the experimental angular
distributions to the simulated ones, we obtained 3$\sigma$
upper limits of the ratio of the E2- to E1-transition
amplitude of the $^7$Be(p,$\gamma$)$^8$B reaction,
$S_{\rm E2}$/$S_{\rm E1}$ of $0.06\times 10^{-4}$,
$0.3\times 10^{-4}$ and $0.6\times 10^{-4}$ for
$E_{\rm rel}=$ 0.3$-$0.5, 0.5$-$0.7 and 1.0$-$1.2 MeV,
respectively. These numbers agree well with the results of Kikuchi
{\it et al.} \cite{Ki97}, and suggest that the results of
the model dependent analysis of Davids {\em et al.} \cite{Davids} needs
to be checked. The MSU collaboration recently also 
published values for $S_{17}(0)$ \cite{Davids2}
which are in agreement with the arlier results of the 
RIKEN II experiment \cite{Ki98}.

\subsection{Conclusion: The Coulomb Dissociation Method}

In conclusion we demonstrated that the Coulomb dissociation (when
used with "sechel") provides
a viable alternative method for measuring small cross section of
interest for nuclear-astrophysics. First results on the CD of \b8
are consistent with
the lower measured values of the cross section and weighted 
average of all thus far data is: $S_{17}(0)\ = \ 18.9 \pm
1.0 \ eV-b$, see Table 1. The accuracy of the extracted S-factors are now
limited by our very understanding of the Coulomb dissociation process,
believed
to be approx. $\pm$10\%. The value of the E2 S-factor as extracted from both
the RIKEN and GSI experiments are consistent and shown to be very small,
$S_{E2}/S_{E1}$ of the order of $10^{-5}$ or smaller, see Table 1.
The adopted value of $S_{17}$ from the CD experiments, shown in Table I 
is in good agreement with recent direct measuremenst \cite{Ha00,Has00}.

\begin{center}

\underline{Table 1:} Measured S-factors in Coulomb 
      dissociation experiments.
\end{center}

\begin{tabbing}

\hspace{1in} \= \underline{Experiment} 
\hspace{1in} \= \underline{$S_{17}(0)$ eV-b}
\hspace{1in} \= \underline{$S_{E2}/S_{E1}$(0.6 MeV)} \\
\   \\
\> RIKEN1 \> $16.9 \pm 3.2$ \> $< 7 \times 10^{-4}$ \\
\   \\
\> RIKEN2 \> $18.9 \pm 1.8$ \> $< 4 \times 10^{-5}$ \\
\    \\
\> GSI1 \> $20.6 \ + 1.2 \ - 1.0$ \> $< 3 \times 10^{-5}$ \\
\    \\
\> GSI2 \>  ?                     \> ?                     \\
\    \\
\> MSU \>  $17.8 \ + 1.4 \ - 1.2$  \> $4.7\ + 2.0\ -1.3 \times 10^{-4} $\\
\    \\
\> \underline{ADOPTED} \> $18.9 \pm 1.0$ \> $< 3 \times 10^{-5}$ \\

\end{tabbing}

\section{Introduction: Oxygen Formation in Helium 
  Burning and The \c12ag Reaction}

The outcome of helium burning is the formation of the two elements, carbon
and oxygen \cite{Fo84,We93,Ga99}.  The ratio of carbon
to oxygen at the end of helium burning
is crucial for understanding the fate of a Type II supernovae 
and the nucleosynthesis of heavy elements. While an oxygen rich star
is predicted to end up as a black hole, a carbon rich star leads to
a neutron star \cite{We93}. At the same time helium burning is also 
very important for understanding Type Ia supernovae 
(SNIa) now used as a standard candle for cosmological distances \cite{SNIa}.
All thus far luminosity calibration curves and the stretching parameter 
are used as empirical observation without fundamental understanding 
between the time characteristics of the light curve and the luminosity 
of a Type Ia supernova.
Since the first burning stage in helium burning, the triple alpha-particle 
capture reaction ($^{8}Be(\alpha,\gamma)^{12}C$), is
well understood \cite{Fo84}, one must extract the p-wave [\sE1] and 
d-wave [\xsE2] cross section of the
$^{12}C(\alpha,\gamma)^{16}O$  reaction at the Gamow peak (300 keV)
with high accuracy of approximately $10\%$  or better to completely 
understand stellar helium burning.
An independent R-matrix analysis
\cite{Ha96} of all available data can not rule out a small
S-factor solution (10-20 keV-b). 
It is thus doubtful that one can indeed rule out
a small E1 S-factor solution based on current data. The 
confusion in this field mandates a direct measurement
of the cross section of the \c12ag reaction at low energies
at energies even .

\section{The Proposed \go16 and $^{16}O(e,e' \alpha)^{12}C$ Experiments}

For determination of the cross section of the \c12ag at very low energies,
as low as $E_{cm}=700$ KeV considerably lower than measured till now
\cite{Hammer01}, it is advantageous 
to have an experimental setup with larger (amplified) 
cross section, high luminosity and low background.  It turns
out that the use of the inverse process, the \go16 reaction may
indeed satisfy all three conditions. The cross section of
\go16 reaction (with polarized photons) at the kinematical region
of interest (photons approx 8-8.5 MeV) is larger by a factor of 
approximately 100 than the cross section of the direct \c12ag reaction. 
Note that the polarization yields
an extra factor of two in the enhancement. Thus for the
lowest data point measured at 0.9 MeV with the direct cross
secion of approx. 60 pb, the photodissociation cross section is 6 nb. It
is evident that with similar luminosities and lower background, see below, 
the photodissociation cross section can be measured to yet 
lower center of mass energies, as low as 0.7 MeV, where the direct \c12ag
crosssection is of the order of 1 pb.
However, we cannot observe cascade gamma decay, 
which are considered to be small at low energies.

The High Intensity Gamma Source (HIGS) \cite{HIGS}, shown in Fig. 2, 
has already achieved many milestones and is rapidly
approaching its design goal of 2-200 MeV gammas, with 9 MeV gammas at a
resolution of 0.1\% and intensity of order $10^9$ /sec. Current 
intensities are of the order of $10^6$ /sec and resolution of 0.8%.
The backscattered photons of the HIGS facility are collimated and 
will enter the target/detector TPC setup as we propose below. 
With a Q value of -7.162, our experiment will utilize
gammas of energies ranging from 8 to 10 MeV.  Note 
that the emitted photons are
linearly polarized \cite{Lit97} and the emitted particles are 
primarily in a horizontal plane with a $sin^2 \phi$ azimuthal 
angular dependence \cite{PRCRC}.  This simplifies the tracking 
of particles in this experiment. The pulsed photon beam provides a 
trigger for the image intensified CCD camera, and the time projection
information from the chamber yields the azimuthal angle
of the event of interest. Background events will be 
discriminated with time of flight techniques, and flushing
of the CCD between two events. To reduce noise, the
CCD will be cooled. We note that similar research program with 
high intensity photon beams and a TPC already exists at the 
Tokyo Institute of Technology \cite{Shima}. An $^{16}O(e,e' \alpha)^{12}C$ 
is proposed at the MIT-Bates accelerator \cite{Genya}. This 
meausrement with virtual photons is sensitive mostly to d-wave 
astrophysical cross section factor.

\centerline{\psfig{figure=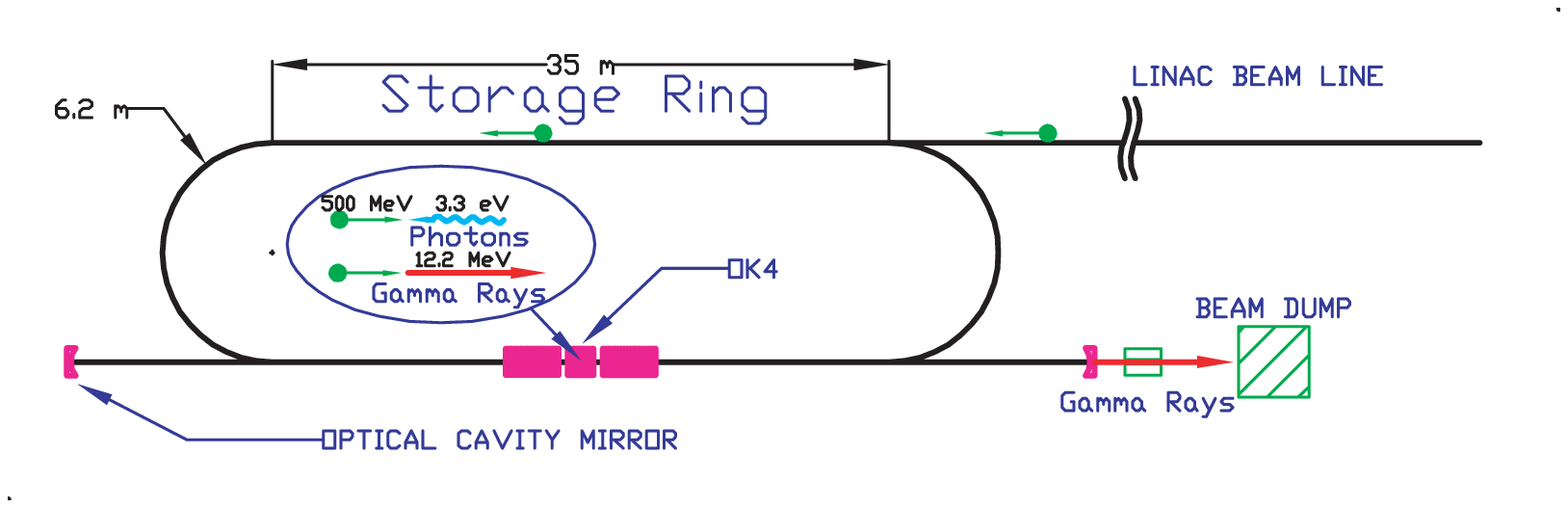,width=6in}}

\hspace{0.5in} \underline{Fig. 2:}  Schematic diagram of the HIGS 
     facility \cite{HIGS} for the production of intense MeV gamma beams.

\subsection{Proposed Time Projection Chamber (TPC)}

We intend to construct an Optical Readout Time Projection Chamber (TPC), 
similar to the TPC constructed in the Physikalisch Technische Bundesanstalt,
(PTB) in Braunschweig, Germany and the Weizmann Institute, Rehovot, 
Israel \cite{NIMA}, for the detection of alphas and carbon, the byproduct 
of the photodissociation of \xo16.  Since the range of available alphas is 
approximately 8 cm (at 100 mbars) the TPC will be 30 cm wide and up to 
one meter long for later phases of our studies. We intend first 
to construct a 30 cm long TPC prototype 
and place it on the HIGS beam line at TUNL/Duke. The TPC that we
intend to construct will be largely insensitive to single Compton electrons,
but will allow us to track both alphas and carbons emitted almost back to back
in time correlation. The very different range of alphas and carbons (approx.
a factor of 4), and differences in the lateral ionization
density, will aid us in particle identification. Such a TPC detector
would also allow us to measure angular distributions with respect to the
polarization vector of the photon thus seperating the E1 and E2 components
of the \c12ag reaction. The excellent energy resolution of the TPC (approx. 
2\%) allows us to exclude events from the photodissociation of nuclei 
other than $^{16}O$, including isotopes of carbon, oxygen and fluorine, which 
may also be present in the gas.  
In Fig. 3, taken from Titt {\em et al.} \cite{NIMA}, we show 
a schematic diagram of the Optical Readout TPC we proposed to construct. A 
photon beam enters the TPC through an entrance hole in the drift chamber part of 
the TPC and mainly produce background $e^+e^-$ pair and a smaller 
amount of compton electrons, as well as the photodissociation 
of various nuclei present in the $CO_2 \ + \ Ar$ gas mixture, including 
$^{16}O$. As we discuss later the background events are easily rejected. 

\centerline{\psfig{figure=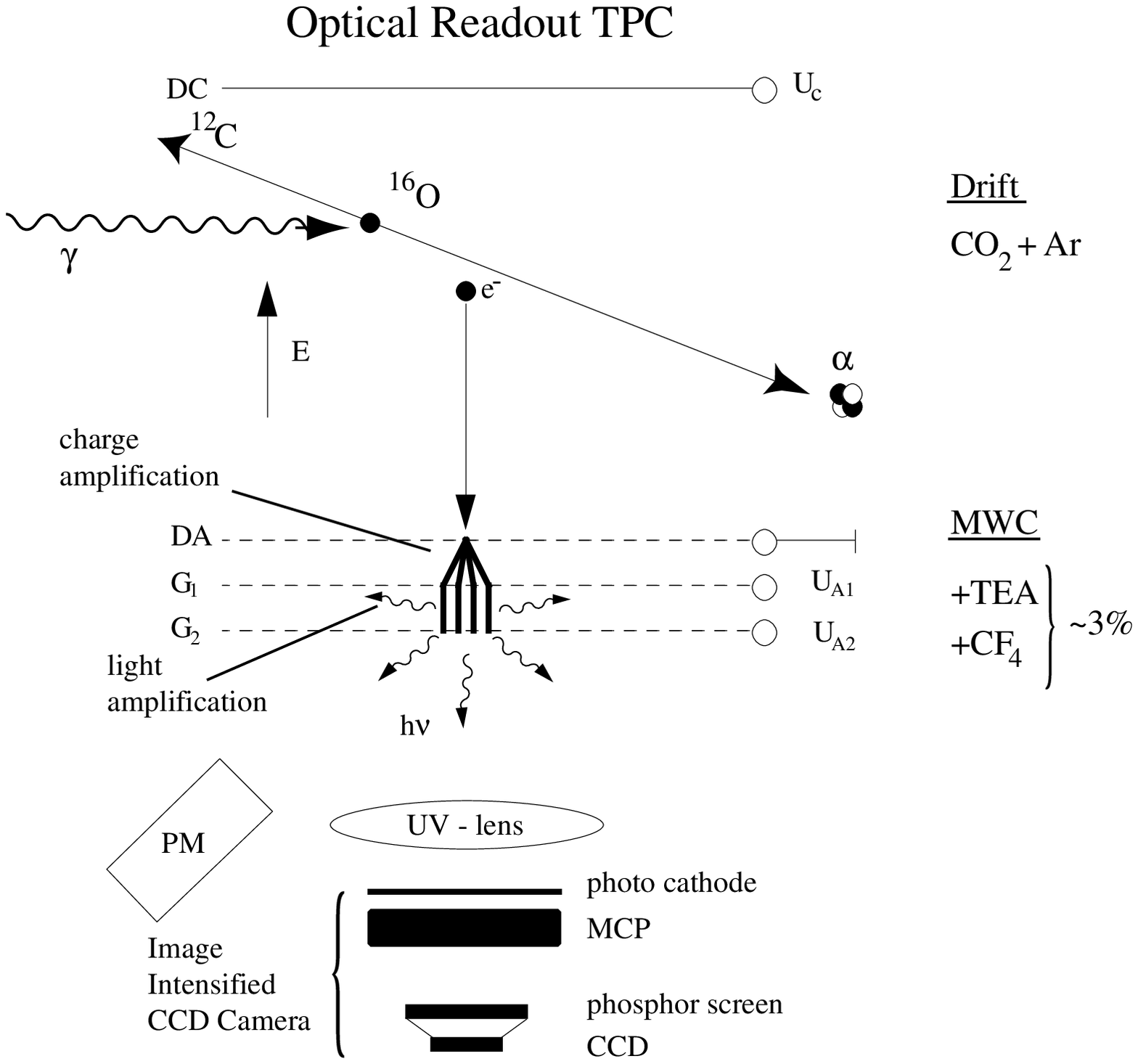,height=2.7in}}

\hspace{0.5in} \underline{Fig. 3:}  Schematic diagram of the Optical
      Readout TPC \cite{NIMA}.

The charged particle byproducts of the 
photodissociation create delta electrons 
that create secondary electrons that drift 
in the chamber electric field with a total time of the order of 1 $\mu s$ 
for 5 cm. The time projection of the drift electrons allows us to measure the 
inclination angle ($\phi$) of the plane of the byproducts, and the tracks themselves 
allow for measurement of the scattering angle ($\theta$), both with an accuracy 
of better than two degrees. The electrons that reach the multi-wire chamber are 
multiplied (by approx. a factor of $10^7$) and interact with a small (3\%) 
admixture of triethylamine (TEA) \cite{NIMA} or $CF_4$ \cite{NIMA2} 
gas to produce UV or visible photons, respectively.
The light detected in the photomultiplier tube, see Fig. 4, triggers the 
Image Intensifier and CCD camera which takes a picture of the visible tracks.
The picture is downloaded to a PC and analyzed for recognition of the two 
back-to-back alpha-carbon tracks. The background electrons lose approx. 100 keV 
in the entire TPC and are removed by an 
appropriate threshold on the Photo Multiplier Tube (PM), and 
events from the photodissociation of nuclei 
other than $^{16}O$ are removed by measuring the total energy (Q-value) of the 
event.

\subsection{Design Goals}

The luminosity of a our proposed \go16 experiment 
can be very large.
For example, with a 100 cm long target of pure $ CO_2 $ at a pressure of 76
torr (100 mbar) and a photon beam of $2 \times 10^9$ /sec,
we obtain a luminosity of
$10^{30}\ sec^{-1}cm^{-2}$, or a day long integrated luminosity of 0.1
pb$^{-1}$.  Thus
a measurement of the photodissociation of $^{16}O$ with cross section of
10 pb, with our proposed high efficiency TPC 
will yield one count per day.  Hence
it is conceivable that a facility with such luminosity and low background
together with the high efficiency TPC will allow us to measure the
photodissociation cross section to a few tens of pb and thus as low as several
hundreds of fb for the direct \c12ag reaction, corresponding to energies as low as 
E$_{cm}$ = 700 keV.

\section{Acknowledgement} 

I would like to acknowledge the work of N. Iwasa, T. Kikuchi,
K. Suemmerer, F. Boue and P. Senger on the data analyses of the 
GSI CD data.  I also acknowledge
discussions and encouragements from
Professors J.N. Bahcall, C.A. Bertulani,
G. Baur, Th. Delbar and H. Weller. The help of Drs. A. Breshkin, 
A. Gandi, and V. Radeka in the design and production of the TPC 
is also acknowledged.

\end{document}